\documentclass[jgrga]{agu2001}
\ifgalley
\let\eject\relax
\fi
\usepackage{agu-ps}
\usepackage{epsfig}
\usepackage[]{graphicx}

\lefthead{Izmodenov et al.}

\righthead{Hydrogen wall and heliospheric absorption}

\journalid{Sometime 2002} \articleid{???}{??}
\paperid{2002???00021} \cpright{AGU}{2002} 

\received{} \revised{} \accepted{1} \published{}

\authoraddr{Vlad Izmodenov,
Lomonosov Moscow State University, Department of Aeromechanics and Gas
  Dynamics, Faculty of Mechanics and Mathematics, Moscow, 119899, Russia
  (izmod@ipmnet.ru)}

\authoraddr{Brian Wood, JILA, University of Colorado, and NIST, Boulder,
  CO 80309-0440 (woodb@casa.colorado.edu)}

\authoraddr{Rosine Lallement,
  Service d'Aeronomie, CNRS, F-91371, Verrieres le Buisson, France
  (Rosine.Lallement@aerov.jussieu.fr)}

\begin{document}

\title{Hydrogen wall and heliosheath Ly-$\alpha$ absorption toward nearby
  stars:  possible constraints on the heliospheric interface plasma flow}

\author{Vlad Izmodenov$^{1}$, Brian E. Wood$^{2}$, Rosine Lallement$^{3}$}
\affil{ (1) Lomonosov Moscow State University, Department of Aeromechanics
  and Gas Dynamics, Faculty of Mathematics and Mechanics, Russia}
\affil{ (2) JILA, University of Colorado, and NIST, Boulder, CO 80309-0440}
\affil{ (3) Service d'Aeronomie, CNRS, F-91371, Verrieres le Buisson, France}

\begin{abstract}
In this paper, we study heliospheric Ly-$\alpha$ absorption
toward nearby stars in different lines of sight.  We use the
Baranov-Malama model of the solar wind interaction with a
two-component (charged component and H atoms) interstellar medium.
Interstellar atoms are described kinetically in the model.  The code
allows us to separate the heliospheric absorption into two components,
produced by H atoms originating in the hydrogen wall and heliosheath
regions, respectively.  We study the sensitivity of the heliospheric
absorption to the assumed interstellar proton and H atom number densities.
These theoretical results are compared with interstellar absorption toward
six nearby stars observed by the Hubble Space Telescope.
\end{abstract}

\begin{article}
\section{Introduction}

The solar system is traveling in the surrounding Local
Interstellar Cloud (LIC).  In the 1960s, it was realized [e.g.,
{\it Patterson et al.}, 1963; {\it Fahr}, 1968; {\it Blum and
Fahr}, 1970] that interstellar atoms penetrate deep into the
heliosphere and, therefore, can be observed.  The interstellar
atoms of hydrogen have been detected by measurements of the solar
backscattered Ly-$\alpha$ irradiance [{\it Bertaux and Blamont},
1971; {\it Thomas and Krassa}, 1971].  Later, interstellar atoms
of helium were also measured directly [{\it Witte et al.}, 1996]
and indirectly as backscattered solar irradiance [{\it Weller and
Meier}, 1981; {\it Dalaudier et al.}, 1984]. At present, there is
no doubt that inside the heliosphere, properties of He atoms such
as temperature and velocity are different from those of H atoms.
In particular, the H atoms are decelerated and heated compared with
atoms of interstellar helium inside the heliosphere [{\it Lallement et
al.}, 1993; {\it Costa et al.}, 1999; {\it Lallement}, 1999]. The
reason for this is a stronger coupling of H atoms with plasma
protons in the heliospheric plasma interface through charge
exchange.  Many works developed the concept of the heliospheric
plasma interface over more than four decades after pioneering
papers by {\it Parker} [1961] and {\it Baranov et al.} [1970].

\begin{figure}
\noindent\includegraphics[width=\hsize]{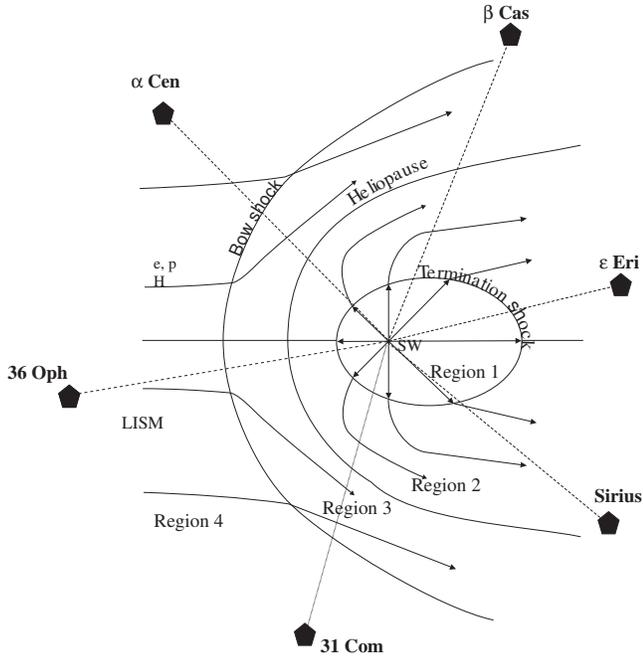}
\caption{ The heliospheric interface is the region of the solar
wind interaction with LIC. The heliopause is a contact
discontinuity, which separates the plasma wind from interstellar
plasmas. The termination shock decelerates the supersonic solar
wind. The bow shock may also exist in the interstellar medium. The
heliospheric interface can be divided into four regions with
significantly different plasma properties: 1) supersonic solar
wind; 2) subsonic solar wind in the region between the heliopause
and termination shock (i.e., the heliosheath) ; 3) disturbed
interstellar plasma region (or "pile-up" region) around the
heliopause; 4) undisturbed interstellar medium. \label{fig1}}
\end{figure}

The heliospheric interface is formed by the interaction of the
solar wind with the charged component of the interstellar medium
(see Figure 1). The heliospheric interface is a complex structure
having a multi-component nature.  The solar wind and interstellar
plasmas, interplanetary and interstellar magnetic fields,
interstellar atoms, Galactic and anomalous cosmic rays (GCRs and
ACRs), and pickup ions all play prominent roles.  Interstellar
hydrogen atoms interact with the heliospheric interface plasma by
charge exchange.  This interaction significantly influences both
the structure of the heliospheric plasma interface and the flow of
the interstellar H atoms.  In the heliospheric interface, atoms
newly created by charge exchange have the properties of local
protons.  Since the plasma properties are different in the four
regions of the heliospheric interface shown in Figure 1, the H
atoms in the heliosphere can be separated into four populations,
each having significantly different properties.  For example,
population 3 consists of the atoms created by charge exchange with
relatively hot protons in the region of disturbed interstellar
plasma around the heliopause (region 3 in Figure 1).  It was
realized theoretically by {\it Baranov et al.} [1991] that the
atoms of population 3 form a so-called ``hydrogen wall'' around
the heliopause. The hydrogen wall is a significant enhancement of
the density of interstellar atoms of hydrogen around the
heliopause compared with the number density of the undisturbed
local interstellar medium. The self-consistent models of the
heliospheric interface predict that the secondary interstellar
atoms (or atoms of population 3) are decelerated and heated
compared with the original interstellar H atoms.

The Ly-$\alpha$ transition of atomic H is the strongest absorption
line in stellar spectra.  Thus, the heated and decelerated atomic
hydrogen within the heliosphere produces a substantial amount of
Ly-$\alpha$ absorption. This absorption, which must be present in
all stellar spectra since all lines of sight go through the
heliosphere, has been unrecognized until recently, because it is
undetectable in the case of distant objects characterized by
extremely broad interstellar Ly-$\alpha$ absorption lines that hide the
heliospheric absorption.  We know now that in the case of nearby
objects with small interstellar column densities, it can be
detected in a number of directions. The heliospheric absorption
will be very broad, thanks to the high temperature of the
heliospheric H, and its centroid will also generally be shifted
away from that of the interstellar absorption due to the deceleration,
allowing its
presence to be detected despite the fact that it remains blended
with the interstellar absorption.  The hydrogen wall absorption
was first detected by {\it Linsky and Wood} [1996] in Ly-$\alpha$
absorption spectra of the very nearby star $\alpha$ Cen taken by
the Goddard High Resolution Spectrograph (GHRS) instrument on
board the Hubble Space Telescope (HST).  Since that time, it has
been realized that the absorption can serve as a remote diagnostic
of the heliospheric interface, and for stars in general, their
``astrospheric'' interfaces.

The $\alpha$ Cen line of sight lies 52$^{\circ}$ from the upwind
direction of the interstellar flow through the heliosphere.  An
additional detection of heliospheric H I absorption only
12$^{\circ}$ from the upwind direction was provided by HST
observations of 36 Oph [{\it Wood et al.}, 2000a].  For downwind
lines of sight, the heliospheric absorption is also not negligible
[Williams et al., 1997]. Analysis of Ly-$\alpha$ absorption toward
Sirius shows that absorption of atoms of population 2 created in
the heliosheath (see Figure~1) is needed in addition to
interstellar and ``hydrogen wall'' absorption components in order
to explain observations [{\it Izmodenov et al.}, 1999].

In addition to heliospheric absorption, ``astrospheric''
absorption towards Sun-like stars was detected by {\it Wood et
al.} [1996], {\it Dring et al.} [1997], and {\it Wood and Linsky}
[1998]. These observations might be used to infer properties of
these stars and their interstellar environments [{\it Wood et al.}
2001; {\it M\"{u}ller et al.}, 2001].

A theoretical model of the heliospheric interface should be
employed to interpret observations and put constraints on the
local interstellar parameters and the heliospheric interface
structure.  This model should self-consistently take into account
plasma and H atom components.  Since the mean free path of H atoms
is comparable with the size of the heliospheric interface, the H
atom flow needs to be treated kinetically through the velocity
distribution function, which is not Maxwellian [{\it Baranov et
al.}, 1998; {\it M\"{u}ller et al.}, 2000; {\it Izmodenov et al.},
2001]. Recently, {\it Wood et al.} [2000b] compared observations of
Ly-$\alpha$ toward six nearby stars with model-predicted
absorption.  Both a Boltzmann mesh code [{\it Lipatov et al.},
1998] and a multi-fluid approach [{\it Zank et al.}, 1996] were
used to compute H atom distributions in the heliosphere.  In
comparing these models with the data, it was found that the
kinetic models predict too much absorption.  Models created
assuming different values of the interstellar temperature and
proton density fail to improve the agreement.  Surprisingly, it
was found that a model that uses a multifluid treatment of the
neutrals rather than the Boltzmann particle code is more
consistent with the data [{\it Wood et al.}, 2000b].

In this paper, we continue to study heliospheric absorption toward
these stars.  In our study, we use the Baranov-Malama model [{\it
Baranov and Malama}, 1993, 1995, 1996] of the heliospheric
interface, which is described in the next section.  This model
uses a Monte Carlo code with splitting of trajectories, which
allows very precise computation of H atom distributions.  Another
advantage of the model is the possibility of separating of
heliospheric H atoms into several populations, as discussed above.
This model advantage allows us to consider separately two types of
heliospheric absorption, hydrogen wall absorption and heliosheath
absorption.

\begin{figure}
\noindent\includegraphics[width=\hsize]{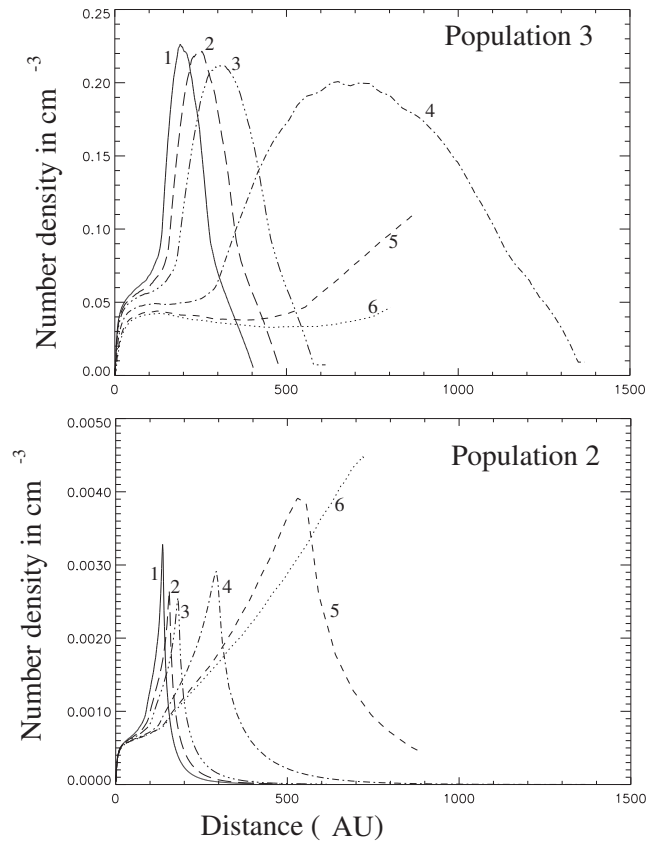}
\caption{Density distributions of H atoms of population 3, atoms
originated in the disturbed interstellar medium (region 3 in
figure 1), and population 2, atoms originated in the heliosheath
(region 2 in figure 2). Density distributions are shown as
functions of heliospheric orientation relative to different line
of sights. Curves 1, 2, 3, 4, 5, and 6 correspond to $\theta =
12^{\circ}$, $52^{\circ}$, $73^{\circ}$, $112^{\circ}$,
$139^{\circ}$, and $148^{\circ}$, respectively, where $\theta$ is
the angle relative to the upwind direction of the interstellar
flow. The figure shows the number densities for model 3 (see Table
1). \label{fig2}}
\end{figure}

\begin{figure}
\noindent\includegraphics[width=\hsize]{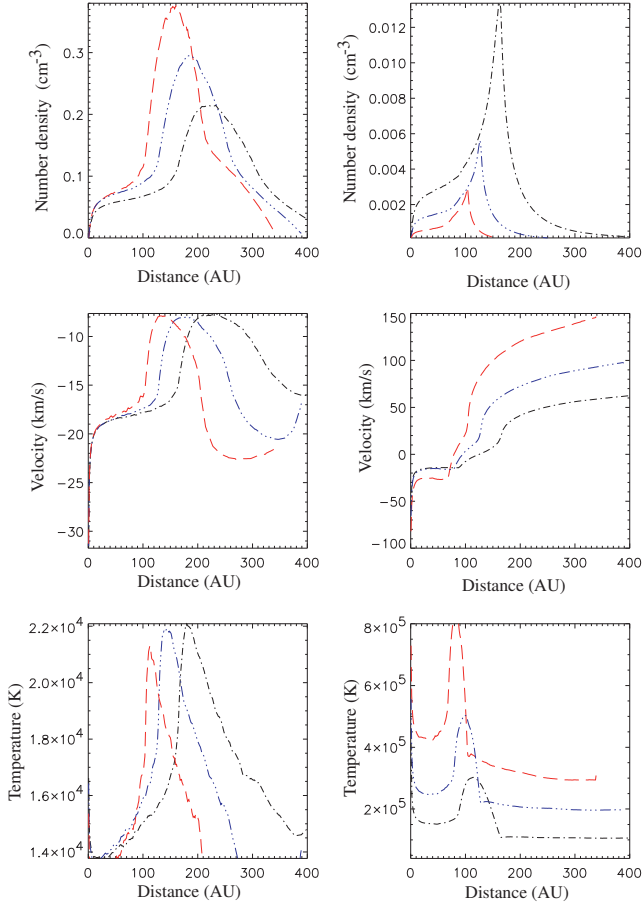}
\caption{ Number densities (upper row), velocities (middle row),
and effective temperatures (lower row) of population 3 (left
column) and population 2 (right column) toward 36 Oph. Dot-dash
lines correspond to model 4, dot-dot-dot-dash lines correspond to
model 5 and long dashes correspond to model 6 (see Table 1).}
\end{figure}

\section{Heliospheric interface model}

To calculate heliospheric absorption we employ the Baranov-Malama
model of the solar wind interaction with the two-component
interstellar medium [{\it Baranov and Malama}, 1993, 1995, 1996;
{\it Izmodenov}, 2000].  This is an axisymmetric model, where the
interstellar wind is assumed to have uniform parallel flow, and
the solar wind is assumed to be spherically symmetric at the
Earth's orbit.  Plasma and neutral components interact mainly by
charge exchange.  However, photoionization, solar gravity, and
solar radiation pressure, which are especially important in the
vicinity of the Sun, are also taken into account in the model. The
process of electron impact ionization, which is important in the
heliosheath [{\it Baranov and Malama}, 1996], is taken into
account as well.  Kinetic and hydrodynamic approaches were used
for the neutral and plasma components, respectively.  The kinetic
equation for the neutrals is solved together with the Euler
equations for a one-fluid plasma.  The influence of the
interstellar neutrals is taken into account in the right-hand side
of the Euler equations that contains source terms [{\it Baranov and
Malama}, 1993; {\it Malama}, 1991], which are integrals of the H
atom distribution function $f_H(\vec{V}_H)$ and can be calculated
directly by a Monte Carlo method [{\it Malama}, 1991]. The set of
kinetic and Euler equations is solved by an iterative procedure
suggested by {\it Baranov et al.} [1991].  Supersonic boundary
conditions were used for the unperturbed interstellar plasma and
for the solar wind plasma at Earth's orbit.  The velocity
distribution of interstellar atoms is assumed to be Maxwellian in
the unperturbed LIC.

Basic results of the model can be briefly summarized as follows
[see also {\it Izmodenov}, 2000].  In the presence of
interstellar H atoms the heliospheric interface is much closer to
the Sun compared with the case of a fully ionized LIC. The
termination shock becomes more spherical, and complicated shock
structure in the tail disappears. The plasma flows are disturbed
upstream of both the bow and termination shocks.  This is an
effect of H atoms that interact with protons by charge exchange.
The solar wind is decelerated by $15-30$\% as it approaches the
termination shock. The Voyager spacecraft observe the deceleration
of the solar wind at large heliocentric distances [{\it
Richardson}, 2001].  The number density of pickup ions, which are
created by charge exchange, may reach $20-50$\% of the solar wind
number density at the distance of the TS.  The expected distance
to the termination shock is $80-100$ AU and depends on local
interstellar parameters.  The interstellar atoms are significantly
disturbed in the interface. Since the velocities of new atoms
created by charge exchange depend on properties of the local
plasma, it is convenient to distinguish four populations of atoms
depending on the place of origin, as described above.  The
velocity distribution function of H atoms can be represented as a
sum of the distribution functions of these populations. {\it
Izmodenov et al.} [2001] studied the evolution of the four
velocity distribution functions in the heliospheric interface.

\begin{table}
\begin{center}
\caption{Sets of model parameters\tablenotemark{a}}
\begin{tabular}{ccccc}
\tableline
Model &  $n_{H,LIC}$ & $n_{p,LIC}$ & Notation in figures  \\
      &  $cm^{-3}$   & $cm^{-3}$   &                      \\
\tableline
1 & 0.10  & 0.10 &  solid      \\
2 & 0.15  & 0.05 &  dotted    \\
3 & 0.15  & 0.10 & dashed     \\
4 & 0.20  & 0.05 & dot-dash    \\
5 & 0.20  & 0.10 & dot-dot-dot-dash    \\
6 & 0.20  & 0.20 & long dash   \\
\tableline
\end{tabular}
\end{center}
\tablenotetext{a}{$n_{H,LIC}$ and $n_{p,LIC}$ are local
interstellar atom and proton number densities, respectively.}
\end{table}


We computed the global heliospheric interface structure and
distribution of interstellar hydrogen in the interface using the
Baranov-Malama model. Our calculations assume the following values
for the fully ionized solar wind at 1 AU:  $n_{p,E} = 6.5$
cm$^{-3}$, $V_{p,E} = 450$ km~s$^{-1}$, and $M_{E}=10$, where
$n_{p,E}$, $V_{p,E}$ and $M_E$ are the proton number density,
solar wind speed, and Mach number, respectively.  For the
inflowing partially ionized interstellar gas, we assume a velocity
of 25.6 km~s$^{-1}$ and a temperature of 7000~K.  These values are
consistent with in situ observations of interstellar helium [{\it
Witte et al.}, 1996]. Two other input parameters, interstellar
proton and H atom number densities, are varied.  The values
assumed for these parameters for the various models are listed in
Table 1.  We vary $n_{p,LIC}$ in the range $0.1 - 0.2$ cm$^{-3}$,
while $n_{H,LIC}$ is varied in the range $0.05 - 0.2$ cm$^{-3}$.

In principle, each of the four H atom populations produces an
absorption line.  However, populations 1 and 4 can be neglected
here, for the following reasons.  Population 4 atoms are
unperturbed interstellar atoms (having not charge exchanged at any
time), and the average temperature and velocity of this population
in the heliosphere is very similar to those of the gas in the
interstellar cloud beyond the heliosphere (there is a slight and
negligible difference due to selection effects).  Thus, the
population 4 absorption cannot be distinguished from the
interstellar absorption.  Since the interstellar column-density is
a fitted parameter, it will automatically include population 4.
Moreover, this gas is not hot and the column density is small,
resulting in an unnoticeable difference.  Population 1 corresponds
to atoms flowing away from the Sun at very large velocities (i.e.,
the speed of the neutralized supersonic solar wind protons).
The produced absorption is thus strongly shifted and broadened due
to solar wind inhomogeneities.  Also, since column densities are
very small, this absorption is very shallow and undetectable.

Figure 2 shows calculated number densities of populations 2 and 3,
which are of interest in this paper, toward six selected line of
sights from upwind to downwind.  The figure shows the number
densities for model 3 (see Table 1).  The hydrogen wall (upper
panel) is most pronounced in the upwind direction.  However, the
wall decreases slowly with increasing of the angle $\theta$
relative to the upwind direction ($\theta=0^{\circ}$ for upwind).
At $\theta = 112^{\circ}$ (curve 4) the wall is still only
slightly lower than it is in the upwind direction.  There is no
hydrogen wall in downwind directions (curves 5 and 6). In upwind
and crosswind directions, the number densities of population 2 are
about two orders of magnitude less than number densities of
population 3. However, the densities of populations 2 and 3 are
different by only about one order of magnitude in the downwind
directions (curves 5 and 6).  Despite their small densities, the
atoms of population 2 can produce noticeable absorption due to
their higher temperature.

Figure 3 shows number densities, bulk velocities, and effective
kinetic temperatures for models 4-6 toward 36 Oph.  The highest
hydrogen wall is for the model with the largest $n_{p,LIC}$ (model
6).  Decrease of $n_{p,LIC}$ results in a decrease of the hydrogen
wall.  At the same time, the hydrogen wall becomes wider, so that
the H column density does not vary significantly from model 4 to
model 6.  All models show approximately the same velocities and
temperatures of population 3 in the hydrogen wall region, which
produces most of the heliospheric absorption.  In contrast to
population 3, the number densities, velocities, and effective
kinetic temperatures of population 2 vary significantly with
$n_{p,LIC}$ (right column plots in Figure 3).  These effects are
connected with: a) the location of the termination shock, which is
closer to the Sun for higher $n_{p,LIC}$, b) the width of the
heliosheath region; and c) effects of interstellar H atoms
filtration and their influence on the heliosheath plasma flow [for
details, see {\it Izmodenov}, 2000; {\it Izmodenov et al.}, 1999,
2001].

More results on the distribution of H atom and plasma parameters
for the presented six models are provided at
http://izmod.ipmnet.ru/~izmod/Papers/IWL/index.html. This page
contains contour plots of plasma and H atom population number
densities and velocities, and other supplemental information.

\section{Modeling of heliospheric absorption}

The absorption profile along a line of sight is
\begin{equation}
I( \lambda) = I_0(\lambda) exp( - \tau(\lambda)),
\end{equation}
where $I_0(\lambda)$ is the assumed background Ly-$\alpha$ profile,
and $\tau(\lambda)$ is the opacity profile.
The opacity profile is
\begin{equation}
\tau = \frac{\pi e^2 f N \lambda^2}{m c^2} \phi(\lambda).
\end{equation}
Here, $f$ is the oscillator absorption strength, $N$ is the column density,
and $\phi(\lambda) = (\lambda_0 c \phi(v)) / \lambda^2 $, where $\phi (v)$
is the normalized velocity distribution along the line of sight.
In the velocity space to which wavelengths can be linearly transferred,
$v = c (1- \lambda_0/\lambda)$, where $\lambda_0$ is the rest wavelength
of Ly-$\alpha$.  In the case of a Maxwellian velocity distribution of the
neutral gas, the absorption is determined by three parameters ($N$, $u$,
$b$), and the absorption profile is
\begin{equation}
I(v) = I_0(v) exp ( -
 \frac{\pi e^2 f N \lambda_0}{m c}
\frac{e^{-\frac{(v-u)^2}{b^2}}}{\sqrt{\pi} b} ).
\end{equation}
Here, $u$ and $b$ are line-of-sight bulk and thermal velocities,
respectively.  Since the column density of H atoms in the
heliospheric interface is less than 10$^{15}$ cm$^{-2}$, the
extended Lorentzian wings of natural line broadening do not
accumulate any appreciable opacity.  Therefore, equation (3),
which corresponds to pure Doppler broadening, is applicable.  Note
that for the interstellar absorption the column density is higher
and one must take the Lorentzian wings into account and replace
the Doppler profile in equation (3) with a Voigt profile:
\begin{equation}
I(v) = I_0(v) exp ( -
 \frac{\pi e^2 f N \lambda_0}{m c \sqrt{\pi} b}
H(a,v-u)).
\end{equation}
Here $a = \Gamma /4 \pi b$, where $\Gamma=0.0755$ is the transition rate in
km/s units, and
\begin{equation}
H(x,y) = \int^{\infty}_{-\infty} \frac{exp(-t^2)
  dt}{(y-t)^2+x^2}.
\end{equation}

Using the distributions of interstellar H atoms computed as
discussed in the previous section, we compute the heliospheric
absorption toward the six selected directions.  To study common
properties and variations of the heliospheric absorption from
upwind to downwind, we assume in this section stellar profiles of
$I_0(\lambda) = 1$.  We compute absorption produced by H atoms of
population 3 and population 2 separately.  Hereafter, these two
absorption components will be called ``hydrogen wall'' and
``heliosheath'' absorption, respectively.

\begin{figure}
\noindent\includegraphics[width=\hsize]{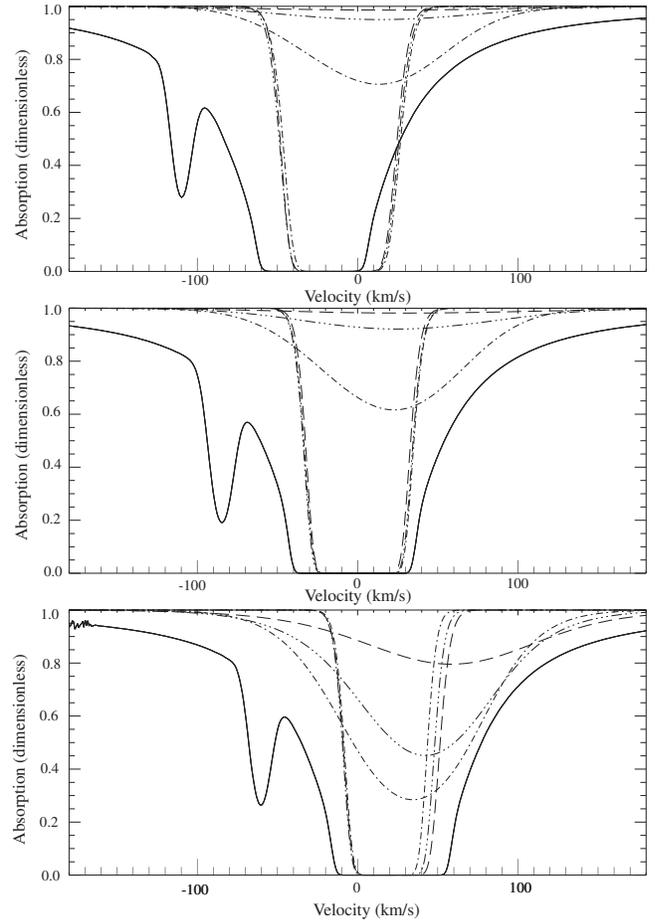}
\caption{ Simulated interstellar and heliospheric absorption of
populations 2 and 3 toward 36 Oph (upper panel), 31 Com (middle
panel), and Sirius (lower panel), which are located in the upwind
(12$^{\circ}$), crosswind (73$^{\circ}$) and downwind
(139$^{\circ}$) directions, respectively. Interstellar absorption
are shown by solid curves. Simulated hydrogen wall and heliosheath
absorption are shown as dot-dash lines for model 4,
dot-dot-dot-dash lines for model 5 and dash lines for model 6 (see
Table 1).  (The hydrogen wall absorption is the narrow, saturated
absorption, while the heliosheath is the broader, unsaturated
absorption.)}
\end{figure}

\begin{figure}
\noindent\includegraphics[width=\hsize]{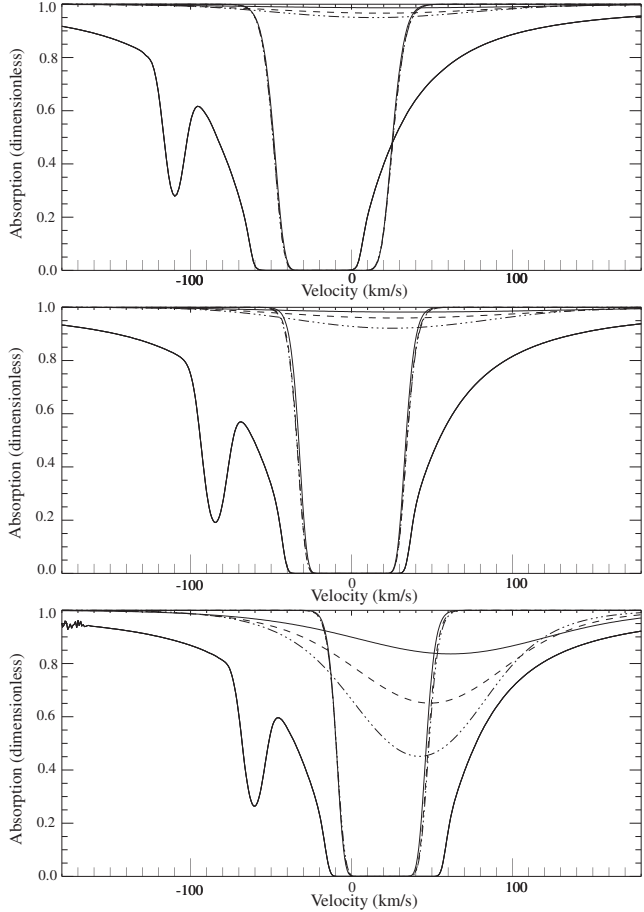}
\caption{ Same as figure 4, but for models 1 (solid lines), 3
(dashed lines), and 5 (dot-dot-dot-dashed lines).}
\end{figure}
Figures 4 and 5 present both calculated hydrogen wall and
heliosheath absorption components for different models of the
heliospheric interface toward three stars, representing nearly
upwind (36 Oph), crosswind (31 Com), and downwind ($\epsilon$ Eri)
directions.  The absorption is shown in the heliospheric rest
frame on a velocity scale.  The expected interstellar
absorption toward these stars is also shown in the figures in
order to compare them with the heliospheric absorption.  Details
of the computation of the absorption are discussed in the next
section.  Figure 4 shows variations of heliospheric absorption
with increase of the interstellar proton number density,
$n_{p,LIC}$, from 0.05 cm$^{-3}$ (model 4) to 0.2 cm$^{-3}$ (model
6), with $n_{H,LIC}=0.2$ cm$^{-3}$ for all these models.  Figure 5
shows variations of heliospheric absorption with an increase of
$n_{H,LIC}$ from 0.10 cm$^{-3}$ (model 1) through 0.15 cm$^{-3}$
(model 3) to 0.2 cm$^{-3}$ (model 5), with $n_{p,LIC}=0.1$
cm$^{-3}$ for all these models.

{\bf Absorption of the hydrogen wall}. All models produce nearly
the same hydrogen wall absorption for upwind and crosswind
directions. Downwind, the absorption is different for different
models.  But due to small column densities of population 3 in
downwind directions, the absorption cannot be detected in stellar
spectra.  In the upwind direction, the atoms of population 3 are
decelerated relative to the original interstellar atoms.
Therefore, the absorption by population 3 is redshifted compared
to the interstellar absorption.  In nearly crosswind directions
the projections of the bulk velocities to the direction of the Sun
are close to zero for population 3 as well as for primary
interstellar gas.  Thus, the absorption of population 3 is hidden
by saturated interstellar absorption and cannot be observed
(middle panels in Figures 4 and 5).

{\bf Absorption in the heliosheath.} In contrast to the hydrogen wall
absorption, the heliosheath absorption varies significantly with
the interstellar proton and H atom number densities.  A higher number
density of population 2 for models with smaller interstellar
proton number density $n_{p,LIC}$ (see Figure 3) leads to large
absorption for this model compared to models with larger
$n_{p,LIC}$.  Models with higher $n_{H,LIC}$ also predict more
heliosheath absorption (see Figure 5) than models with small
$n_{H,LIC}$.  The difference between models
increases for crosswind and downwind
directions.  Generally, for all models the heliosheath absorption
is more pronounced in the crosswind and downwind directions.
Toward these directions both the size of the heliosheath and
the number density of population 2 are larger compared with upwind.
The heliosheath absorption is redshifted in crosswind directions
compared with the interstellar and hydrogen wall absorption components,
and analysis of absorption toward crosswind stars might put
particularly strong constraints on the heliosheath plasma
structure.

As seen from Figures 2 and 4, the parameters of heliospheric
populations of H atoms have large variations in the heliospheric
interface, as well as in each of its sub-regions.  Moreover, the
velocity distributions of the heliospheric populations are not
Maxwellian [{\it Izmodenov et al.}, 2000].  However, it is still
interesting to see whether it is possible to model heliospheric
absorption as if it were produced by uniform Maxwellian gases.
Using the computed distribution functions, we
take the zeroth, first, and second moments of the Maxwellian to
compute $N$, $u$, and $b$. Then, we use
these numbers to compute a Gaussian profile according to eq. (3).
A comparison of the original profiles computed directly from the
velocity distributions with the Gaussian profiles shows rather
good agreement for both populations 2 and 3 (see Figure 6).  Note
that we compute Gaussian profiles and compare them with the exact
profiles for populations 2 and 3 separately.  If populations 2 and
3 were treated as one population, the Gaussian approximation would
be much worse.  Note that for some directions and models the
difference between model predictions and their Gaussian
approximations might be larger, and these fits should be used for
interpretation with great caution.

It is important to note here that our model does not extend far
enough downwind.  Our computational grid extends only 700 AU in
the downwind direction, which is not far enough to incorporate all
heliospheric absorption [{\it Izmodenov and Alexashov}, in
preparation]. Therefore, our calculations underestimate absorption
in downwind directions ($\theta > 125^{\circ}$).

\begin{table}
\begin{center}
\caption{Information on stars and interstellar
absorption.\label{tbl-2}}
\begin{tabular}{ccccccc}
\tableline\tableline Star &  Angle  & Distance  &
  \multicolumn{4}{c}{Interstellar parameters toward the star\tablenotemark{a}}  \\
    &    from          &    (pc)              & $u$          &  T    &  N
                              & D/H,                     \\
    &    Upwind                 &                  & (km s$^{-1}$)  &     (K) &
$\times$ 10$^{17}$ &      $\times$ 10$^{-5}$  \\
    &                  &                  &  &      &
(cm$^{-2}$) &      \\
\tableline
36 Oph        & $12^{\circ}$& 5.5   & -28.4 & 6200  & 7.80  &1.25   \\
$\alpha$ Cen  & $52^{\circ}$& 1.34  & -18   & 5800  & 4.00  &1.5   \\
31 Com        & $73^{\circ}$&90.9   & -3.2  & 8000  & 8.3   &1.75  \\
$\beta$ Cas   &$112^{\circ}$&13.9   &  9.1  & 9500  &13.5   &1.65   \\
 Sirius       &$139^{\circ}$& 2.7   & 19    & 6000  & 1.6   &1.65  \\
              &             &       & 13    & 6000  & 1.6   &1.65  \\
$\epsilon$ Eri&$148^{\circ}$& 3.3   & 20.8  & 7300  & 7.6   &1.45   \\
\tableline
\end{tabular}
\end{center}
\tablenotetext{a}{$u$,T,N are the interstellar velocity,
temperature, column density and D/H ratio toward the star.}
\end{table}

\section{The observations, interstellar absorption, and stellar spectra}

In this section, we use the heliospheric absorption models
discussed above to analyze absorption spectra toward six nearby
stars. For accurate study of the heliospheric absorption based on
the analysis of observed absorption spectra, one needs to know
both the original stellar Ly-$\alpha$ line profile and the
interstellar absorption.

The Ly-$\alpha$ lines of these six stars observed by HST are shown
in Figure 7 (left column).  All of the spectra except that of 36
Oph A were taken using the Goddard High Resolution Spectrograph
(GHRS) instrument.  The 36 Oph A data were obtained by the Space
Telescope Imaging Spectrograph (STIS), which replaced GHRS in
1997.  The spectra are plotted on a heliocentric velocity scale.
All show broad, saturated H I absorption near line center and
narrower deuterium (D I) absorption about 80 km s$^{-1}$ blueward
of the H I absorption.  Figure 7 also shows the assumed intrinsic
stellar Ly $\alpha$ lines and the best estimates for interstellar
absorption.  The interstellar parameters of these estimates are
summarized in Table 2.  The models suggest that heliospheric absorption
will always be redshifted from the the ISM absorption regardless of
the direction of the line of sight, while astrospheric absorption will
conversely be blueshifted [{\it Wood et al.}, 2000b; {\it Izmodenov et al.}
1999].  This is primarily because of the deceleration and deflection of
neutral H at the bow shock, and because for the heliosphere we are
observing the decelerated and deflected neutral H from inside the
heliosphere while for astrospheres
we are observing it from the opposite perspective outside the astrosphere.
Thus, excess absorption on the red side of the
Ly $\alpha$ absorption, when present, is interpreted as
heliospheric absorption, while excess absorption on the blue side,
when present, is interpreted as astrospheric absorption.
In this paper we focus on the red side of the Ly
$\alpha$ absorption since we are interested in the heliospheric
absorption. The stellar profiles and
interstellar absorption estimates are based on previously
published work, which is summarized below.

{\bf $\alpha$ Cen.} The $\alpha$ Cen data provided the first
evidence for heliospheric absorption.  Both members of the
$\alpha$ Cen binary system were observed.  {\it Linsky and Wood}
[1996] demonstrated that the results of their analysis were the
same for both the $\alpha$ Cen A and $\alpha$ Cen B data.  It is
the $\alpha$ Cen B spectrum that we display in Figure 7, since it
has a somewhat higher signal-to-noise ratio (S/N) than the
$\alpha$ Cen A data.  Using the D I line to define the central
velocity and temperature of the interstellar material, {\it Linsky
and Wood} [1996] found they could not fit the $\alpha$ Cen H I
profile with only interstellar absorption.  In Figure 7, we show
interstellar absorption toward $\alpha$ Cen with the D/H value
equal to $1.5\times 10^{-5}$, which is the generally accepted LIC
D/H value.

{\bf 36 Oph.} {\it Wood et al.} [2000a] analyzed the 36 Oph data.
The analysis of the 36 Oph data proved to be very similar to that
of $\alpha$ Cen, with the apparent existence of excess Ly-$\alpha$
absorption on both the red and blue sides of the line.  {\it Wood
et al.} [2000a] argue that only contributions of both heliospheric
and astrospheric absorption can account for all of the excess
absorption.

{\bf Sirius} The Sirius data were first presented by {\it Bertin
et al.} [ 1995a, 1995b]. Unlike the other lines of sight presented
here, there are two interstellar components instead of just one.
Nevertheless, {\it Bertin et al.} [1995a] found that if they
constrained the H I absorption by assuming the LISM temperature
and D/H ratio reported by {\it Linsky et al.} [1993] toward
Capella (T = 7000 K and D/H = $1.65\times 10^{-5}$), they could
not account for excess H I absorption on both the blue and red
sides of the line. {\it Izmodenov et al.} [1999] proposed that the
red excess is due to heliospheric absorption and the blue excess
is due to astrospheric absorption.

{\bf 31 Com, $\beta$ Cas and $\epsilon$ Eri.} The 31 Com, $\beta$
Cas and $\epsilon$ Eri data were first analyzed by {\it Dring et
al.} [1997].  The 31 Com and $\beta$ Cas spectra can be fitted by
single interstellar absorption components, but the $\epsilon$ Eri
spectrum is more complex.  The H I absorption is significantly
blueshifted relative to the D I absorption, indicating a
substantial amount of excess H I absorption on the blue side of
the line.  {\it Dring et al.} [1997] interpreted this excess
absorption to be astrospheric in origin.

\begin{figure}
\noindent\includegraphics[width=6cm]{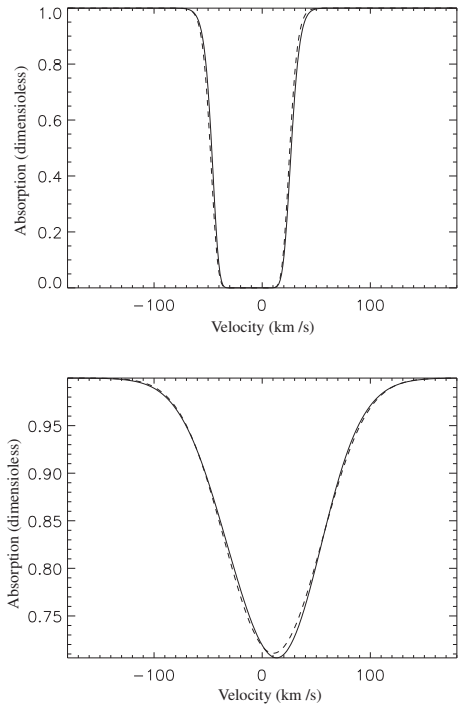}
\caption{A comparision of absorption profiles calculated directly
from distributions of H atoms (solid lines) with absorption
profile approximations assuming uniform Maxwellian gases (dashed
lines). The absorption profiles are shown toward the 36 Oph
direction ($\theta=12^{\circ}$). Upper and bottom panels are for
populations 2 and 3, respectively.}
\end{figure}

\nonewpage

\begin{figure*}
 \noindent\includegraphics[width=15.5cm, clip]{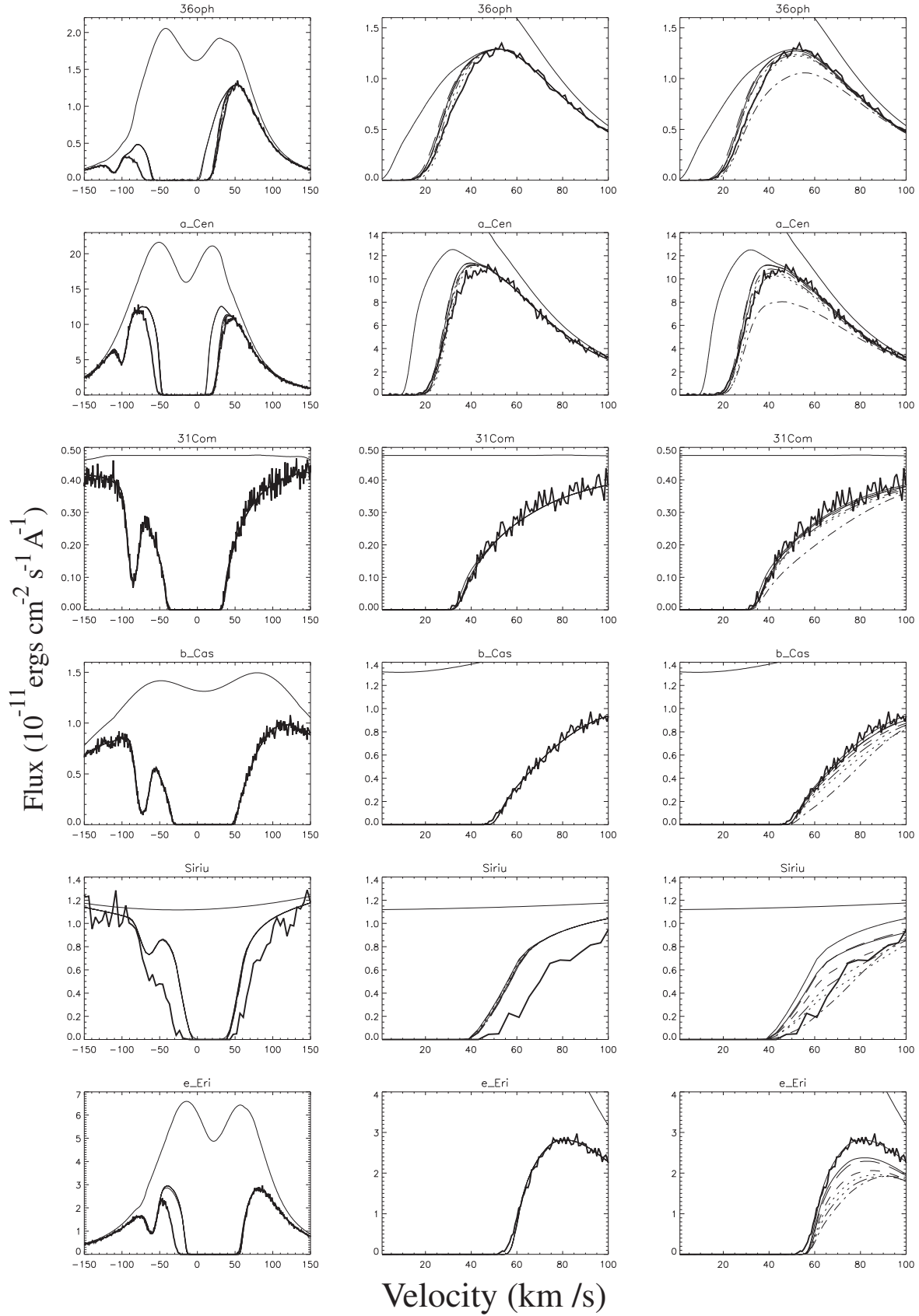}
 \caption{Left column: HST Ly $\alpha$ spectra of six
stars: 36 Oph, $\alpha$ Cen, 31 Com, $\beta$ Cas, Sirius,
$\epsilon$ Eri, respectively from top to bottom. Each plot shows
the observed profile (thick solid line), assumed stellar line
profile and interstellar absorption (thin solid lines). Middle
column: Reproduction of left column, zoomed in on the red side of
Ly-$\alpha$ absorption line. In addition, simulated Ly-$\alpha$
profiles after interstellar and hydrogen wall absorption are shown
for models 1-6. Solid lines, when it is different from assumed
stellar profile and interstellar absorption, correspond to model
1; dotted lines show results of model 2; dashes correspond to
model 3; dot-dash curves correspond to model 4, dot-dot-dot-dash
lines correspond to model 5, long dashes correspond to model 6.
Right column: same as the middle column, but heliosheath
absorption is added to the simulated Ly-$\alpha$ profiles.}
\end{figure*}

\section{Comparison of theory and observations}

The middle and right columns of Figure 7 show the Ly $\alpha$
lines zoomed in on the red side of the HI absorption line, since
that is where most of the heliospheric absorption is expected. The
Ly $\alpha$ profile after interstellar absorption is shown in the
plots as solid lines.  The middle column also shows Ly $\alpha$
profiles after interstellar and hydrogen wall absorption, while
the right column shows Ly $\alpha$ profiles after interstellar,
hydrogen wall, and heliosheath absorption. The hydrogen wall and
heliosheath components are shown for all six models discussed
above (see Table 1).

The absorption spectra toward upwind directions (36 Oph and
$\alpha$ Cen) are fitted rather well by hydrogen wall absorption
only. In comparing the theoretical absorption with data, the most
important place for the model to agree with the data is at the
base of the absorption, where it is very difficult (if not
impossible) to correct any discrepancies by altering the assumed
stellar Ly $\alpha$ profile [{\it Wood et al.}, 2000b].  Toward 36
Oph, model 2 produces slightly more absorption and model 6
produces slightly less absorption at its base compared to the
data.  For other models the differences between model predictions
and HST data at the base can be reduced by small corrections to
the assumed stellar profile and interstellar absorption.  Toward
$\alpha$ Cen, all models show slightly more absorption at the base
than the data, except for model 6.  It is interesting to note that
model 6 fits the $\alpha$ Cen data remarkably well, while it does
not fit as well the Ly $\alpha$ spectrum of 36 Oph.  Note that
these discrepancies with the data just discussed are small enough
that they could in principle be corrected by reasonably small
changes to the assumed stellar Ly $\alpha$ profile or interstellar
absorption, although the better fits in Figure 7 are still to be
preferred.

The heliosheath absorption, which is added to the interstellar and
hydrogen wall absorption in the right column of Figure 7, does not
change the fits much for the upwind lines of sight. Model 2
predicts too much absorption toward both 36 Oph and $\alpha$ Cen
and produces the greatest difference with the data at the base of
the absorption where it matters most.  Consideration of the
heliosheath absorption significantly increases the total
heliospheric absorption predicted by model 4.  The models with the
worst agreement in the upwind directions (model 2 and model 4)
correspond to the extreme case of small interstellar proton number
density, $n_{p,LIC}$.

The observed absorption in the crosswind lines of sight to 31 Com and
$\beta$ Cas shows no need for any heliospheric absorption, and our models do
not generally predict significant absorption in the crosswind directions
that can be clearly distinguished from interstellar absorption.
Significant heliosheath absorption in model 4 makes the model the
worst fit to the data in the crosswind directions.  The other models
fit the data acceptably well.

Results of the comparison for downwind lines of sight are more
puzzling.  It is clearly seen that the missing absorption toward
Sirius can be easily explained by absorption in the heliosheath.
Moreover, the heliosheath absorption predicted by different models
is noticeably different.  Model 4, which has the
worst fit to the data for all other lines of sight, fits the
observed Sirius spectrum better than other models.

In contrast to Sirius, for $\epsilon$ Eri the interstellar
absorption fits the observed spectrum very well and there is no
need for heliospheric absorption.  There is no hydrogen wall
absorption in this direction (middle column of Figure 7), but all
of our models predict too much heliosheath absorption.  A similar
problem was also found by {\it Wood et al.} [2000b] when they
compared their kinetic models with the data.  The discrepancy may be even
more dramatic than our models suggest, because our computational
grid extends only 700 AU in the downwind direction, which as
stated above is not far enough to account for all the absorption.
A larger grid would presumably make the $\epsilon$ Eri
discrepancies worse for all models, and also change which model
fits best for Sirius.

\section{Discussion}

As seen from the previous section, absorption spectra toward
five of six stars can be explained by taking into account the
hydrogen wall absorption only.  Only the Sirius line of sight requires
a detectable amount of heliosheath absorption to fit the data.
Unfortunately, the hydrogen wall absorption is not very sensitive to
such interstellar parameters as the interstellar proton and H atom number
densities.  The hydrogen wall absorption is most apparent and detectable
in upwind directions.  The small differences between the models and
observations in these directions can be eliminated by small
alterations of the stellar profiles (see Figure 7).  The poorest
agreement is for model 2.  After including heliosheath
absorption, model 4 seems to predict too much heliosheath
absorption compared with the upwind data.  Therefore, model 2 and model 4
produce the worst agreement with the data.  This conclusion is also
consistent with the analysis of absorption in crosswind directions (31 Com,
$\beta$ Cas).

Downwind lines of sight (Sirius and $\epsilon$ Eri) can serve as
good diagnostics of the heliosheath absorption.  For the
assumed stellar line profiles, all models predict too much absorption
toward the most downwind line of sight, $\epsilon$ Eri.  However, since
most of the discrepancy is away from the base of the absorption, we can
try to correct this problem by modifying the shape of the assumed
stellar Ly-$\alpha$ profile.  Figure 8 demonstrates how the
stellar profile of $\epsilon$ Eri would have to be modified in order to
improve agreement with the data.  We do not display the central part of
the profile, because the absorption is saturated, providing no
information on the stellar Ly-$\alpha$ profile there.  Because the models
predict too much absorption, the fluxes of the assumed stellar profile
must be increased to improve the fit.

The revised profiles in Figure 8 are plausible in that they
generally do not contain extremely steep slopes or fine structure.
However, it is questionable if the tallest of these profiles (for
models 4 and 5) are truly realistic. The revised stellar profiles
in Figure 8 would suggest that the stellar Ly $\alpha$ profile of
$\epsilon$ Eri is significantly taller and narrower than the line
profiles of similar stars like $\alpha$ Cen B and 36 Oph (see
Figure 7), and the Sun.  There is some subjectivity in deciding
what is plausible and what is not, but we would conclude from
Figure 8 that models 4 and 5 probably require unrealistic
modification to the stellar line profile to fit the data, but the
other four models that require less extreme alterations of the
assumed stellar profile may be all right.  In truth, however, the
situation is actually worse than this, because as mentioned above
our models do not extend far enough downwind and therefore
underestimate the amount of heliosheath absorption.  Since a
larger grid size would significantly worsen agreement with the
data, it is uncertain whether any of the six models is truly
consistent with the $\epsilon$ Eri data.  Furthermore, since the
kinetic models presented by {\it Wood et al.} [2000b] had the same
problem with the $\epsilon$ Eri line of sight, it is possible that
inaccuracies in the physical assumptions involved in current
kinetic models may be resulting in overpredictions of absorption
in downwind directions, although additional theoretical work is
required to test this interpretation.

The models that have the least disagreement with the $\epsilon$ Eri
line of sight are models 1 and 6.  The common feature of these models is
that both predict small number densities of H atoms of population 2.  In
model 1 this is due to a low assumed interstellar H density, while in
model 6 this is due to the small size of the heliosheath.  Model 6 in fact
predicts the narrowest heliosheath region of our six models.  However,
the interstellar proton and H atom number densities of this model
seem to be unrealistically high, because the model predicts that the
termination shock should be at 70 AU in the upwind direction.  As of August
2001, Voyager 1 was at 82 AU and Voyager 2 was at $\sim$65 AU, and
neither has yet encountered the termination shock.  Due to its low assumed
interstellar H density, model 1 predicts the smallest amount of H atoms
penetrating through the heliopause into the heliosphere of our six models.
The number density of H atoms at the termination shock is 0.056 cm$^{-3}$
for this model.  This value is very close to the number density derived
from the observed deceleration of the solar wind due to interaction with
interstellar H atoms.  Comparison of the solar wind speed measured on
Ulysses and Voyager shows a decrease of about 40 km~s$^{-1}$, or 10\% in
radial speed near 60 AU.  Wang and Richardson [2001] have shown that
this speed decrease implies an interstellar neutral density at the
termination shock of 0.05 cm$^{-3}$, which is close to our model 1
value.

Consideration of the Sirius line of sight complicates matters further.
Model 2 and especially model 4 fit the observed profile very well,
once again focusing our attention at the base of the absorption
(see Figure 7), while the other models predict too little
absorption. Models 2 and 4 assume a small interstellar ionization
fraction in the vicinity of Sun, and recent studies of the LIC
ionization state are in favor of such small ionization [{\it
Lallement}, 1999]. However, model 4 does not fit the other lines
of sight well, and it must be noted once again that the models
will underestimate the amount of heliosheath absorption in
downwind directions due to the limited extent of the current model
grid.

The Ly-$\alpha$ absorption profile of the Sirius spectrum has been
a controversial subject for the last few years and will probably
remain so.  First interpreted as due to absorption by a stellar
wind from Sirius A [{\it Bertin et al.}, 1995a], the extra
absorption on the blue side of the line has later been explained
as resulting from an astrosphere around Sirius and shown to be
compatible with a very crude model of such an astrosphere.
However, the comparison between the Sirius B and Sirius A spectra
later obtained by {\it Hebrard et al.} [1999] has shown that a
large fraction (if not the totality) of the extra absorption is
specific to Sirius A and must arise closer to the star than
predicted by an astrospheric model, favoring again a stellar wind
as the source of the absorption.  On the other hand, the extra
absorption on the red side, first interpreted as due to
interstellar hot gas [{\it Bertin et al.}, 1995b], has later been
compared satisfactorily with heliospheric absorption by the
heliosheath [{\it Izmodenov et al.}, 1999], an interpretation
which appears plausible based on Figure 7.  Unfortunately, the
comparison with Sirius B does not help here to disentangle the
sources of the absorption, since the Sirius B spectrum shows a
broader and deeper absorption on the red wing, which encompasses
the Sirius A line, and is identified by {\it Hebrard et al.}
[1999] as due to the intrinsic photospheric Ly-$\alpha$ profile.
{\it Hebrard et al.} [1999] also note that the excess absorption
on the red side of the line can be removed if one relaxes the
assumption that both interstellar components detected toward
Sirius have ${\rm D/H}=1.6\times 10^{-5}$.  The heliosheath or
interstellar gas remain possible sources for the excess absorption
seen towards Sirius A on the red side of the line.  If the
absorption is due to the heliosheath, this is the only line of
sight in which heliosheath absorption is detected.

We now suggest one possible cause of our difficulties with
downwind lines of sight.  The process of pickup ion assimilation
into the solar wind is a very complicated phenomenon.  Currently,
it is believed that pickup ions form a separate population of
protons, which is co-moving with the solar wind proton plasma. The
process of energy exchange between the populations is quite slow
and it is expected that the relaxation time is large compared with
the time that is needed for a proton to reach the termination
shock distance.  How the non-equilibrium distribution of pickup
ions is affected by the termination shock is still an open
question, although some scenarios have been developed [e.g., {\it
Fichtner}, 2001]. Our model considers the solar protons and pickup
ion protons as one fluid.  This approach is based on fundamental
questions of mass, momentum, and energy conservation. Therefore,
it is quite natural to expect that the model predicts the
positions of the shocks and the heliopause reasonably well. At the
same time, measurements of speed and temperature of the solar wind
at large heliocentric distances by Voyager are in favor of the two
populations being co-moving, but thermally different. This
suggests a different scenario of the heliosheath plasma flow than
is assumed in our model.

This new scenario would result in different distributions of H
atoms created in the heliosheath, which could possibly solve our
difficulties in modeling the heliospheric absorption in downwind
directions.  {\it A priori}, we cannot exclude a scenario that
provides the greatest absorption at heliocentric angles of about
$130-150^{\circ}$ from upwind.  For such a scenario, perhaps all
six lines of sight in Figure 7 could be explained simultaneously.
Calculating energetic neutral atom (ENAs) fluxes, {\it Gruntman et
al.} [2001] consider different plasma scenarios of heliosheath
plasma flow.  One of the models considered in the paper suggests a
maximum of ENA fluxes in directions of $130-150^{\circ}$ from
upwind.  This model assumes that the pickup proton population is
carried through the shock without thermalization, i.e., preserving
its velocity distribution filling the sphere in velocity
space.  Since ENAs originate in the heliosheath, it supports the
idea to consider different scenarios of the heliosheath plasma in
order to interpret all six observable spectra simultaneously.  It
remains to be seen whether such models could both reproduce the
absorption seen toward Sirius and solve our problem of
overpredicting absorption toward $\epsilon$~Eri.

Finally, we have to note that an axisymmetric model of the
heliospheric interface was used to calculate the heliospheric
absorption. The actual heliosphere is not axisymmetric due to both
the latitudinal variation of the solar wind and possible effects
of the interstellar magnetic field.  Note, however, that the
upwind direction of the ISM flow is only about $8^{\circ}$ from the
ecliptic, so variations of solar wind properties with latitude are
unlikely to produce major differences from the axisymmetric
approximation.  In any case, asymmetry in the heliosheath
or hydrogen wall could result in the heliospheric absorption being
different from axisymmetric.  A 3D model is needed to assess the
inaccuracies introduced by the axisymmetric assumption, but unfortunately
there is as yet no adequate 3D model including the interstellar H
atom population that can address this issue.
Such a model is currently being developed by us.

At the same time, our results can be interpreted as indirect
evidence that there is no strong deviation of the heliospheric
interface from being axisymmetric,
since all four upwind/crosswind lines of sight discussed here
can be well interpreted on the base of axisymmetric model.
Taking into account the discussion above, one may conclude that
our difficulties with the downwind lines of sight have
other interpretations. However, this
statement needs to be verified by 3D modeling of the heliospheric
interface and by studying the sensitivity of heliospheric absorption
to the 3D structure.

\begin{figure}
\noindent\includegraphics[width=\hsize]{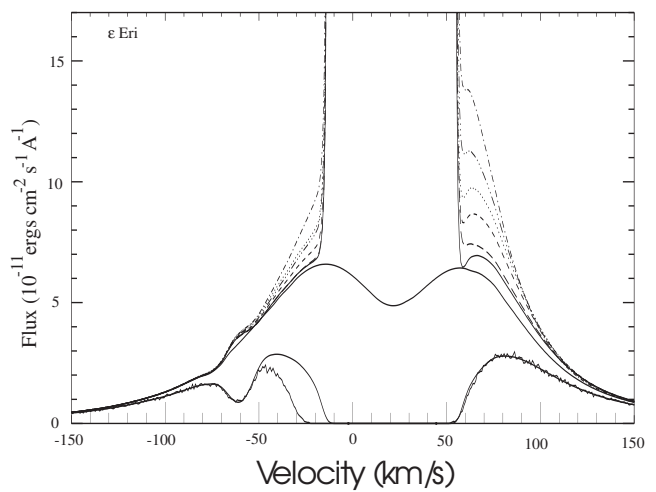}
\caption{Modifications to the assumed stellar Ly-$\alpha$ profile
of $\epsilon$ Eri that lead to better fits to the data for our 6
models. The lines are identified with the models the same as in
Figure 7.}
\end{figure}

\section{Summary}

We have compared H I Ly-$\alpha$ absorption profiles toward six nearby
stars observed by HST with theoretical profiles computed using Baranov-Malama
model of the heliospheric interface with six different sets of model
parameters.  Our results are summarized as follows:

1. It has been shown that the absorption produced by the hydrogen
wall does not depend significantly on local interstellar H atom
and proton number densities for upwind and crosswind directions.
In downwind directions the hydrogen wall absorption is
sensitive to interstellar densities, but this absorption component is
most easily detected in upwind directions.  In crosswind and
downwind directions the hydrogen wall absorption is hidden in the
saturated interstellar absorption and cannot be observed.

2. The heliosheath absorption varies significantly with
interstellar proton and H atom number densities.  For all models,
the heliosheath absorption is more pronounced in crosswind and
downwind directions.  The heliosheath absorption is redshifted in
crosswind directions compared with the interstellar and hydrogen wall
absorption components.

3. Comparison of computations and data shows that all available
absorption spectra, except that of Sirius, can be explained
by taking into account the hydrogen wall absorption only.
Considering heliosheath absorption, we find that all models
have a tendency to overpredict heliosheath absorption in
downwind directions.  Toward upwind and crosswind stars the small
differences between model predictions and the data can be corrected
by small alterations of the assumed stellar Ly-$\alpha$ profile.
However, the downwind $\epsilon$ Eri line of sight is a problem,
as the models predict too much heliosheath absorption in that
direction, and for many, if not most, of the models the discrepancy
with the data is too great to resolve by reasonable alterations of
the stellar profile.

4. It is puzzling that model 4 provides the best fit to the
absorption profile toward Sirius but the worst fit to the other lines
of sight.  This may be due to our models underestimating heliosheath
absorption in downwind directions due to limited grid size, or perhaps
the detected excess absorption toward Sirius is not really heliospheric
in origin, as suggested by previous authors.  It is also possible that
the difficulties the models have with the downwind lines of sight
towards Sirius and $\epsilon$ Eri might be resolved by modifications
to the models, perhaps by taking into account the multi-component
nature of the heliosheath plasma flow.

\acknowledgments This work was supported in part by INTAS Awards
2001-0270, YSF 00-163, RFBR grants 01-02-17551, 02-02-06011,
01-01-00759, CRDF Award RP1-2248,  and International Space Science
Institute in Bern. We thank Horst Fichtner and another referee for
useful suggestions.

\end{article}

\end{document}